\documentclass[twocolumn,twoside,slac_two]{revtex4}
\usepackage{graphicx}
\usepackage{fancyhdr}
\usepackage{natbib}
\newcommand{\fermilat}{\textit{Fermi}-LAT} 

\begin{document}

\title{Very high energy gamma-ray follow-up observations of novae and dwarf novae with the MAGIC telescopes.}


\author{J. Sitarek}
\affiliation{University of \L\'od\'z, PL-90236 Lodz, Poland, \\
IFAE, Campus UAB, E-08193 Bellaterra, Spain}
\author{W. Bednarek}
\affiliation{University of \L\'od\'z, PL-90236 Lodz, Poland}
\author{R.~L\'opez-Coto}
\affiliation{IFAE, Campus UAB, E-08193 Bellaterra, Spain}
\author{E.~de O\~na Wilhelmi for the MAGIC Collaboration}
\affiliation{Institute of Space Sciences, E-08193 Barcelona, Spain}

\author{R. Desiante}
\affiliation{Universit\`a di Udine, and INFN Trieste, I-33100 Udine, Italy}

\author{F. Longo}
\affiliation{Universit\`a di Trieste and INFN Trieste, Italy}

\author{E. Hays for the \fermilat\ Collaboration}
\affiliation{NASA Goddard Space Flight Center, Greenbelt, MD 20771, USA}

\begin{abstract}
In the last few years the \fermilat\ instrument has detected GeV gamma-ray emission from a few novae. 
Such GeV emission can be interpreted in terms of an inverse Compton process of electrons accelerated in a shock. 
It is expected that hadrons can be accelerated in the same conditions, but reaching much higher energies. 
They can produce a second component in the gamma-ray spectrum at TeV energies. 
We performed follow-up observations of selected novae and dwarf novae in a search of the second component in the gamma-ray spectrum. 
This can shed light on the acceleration process of leptons and hadrons in nova
explosions. 
We have performed observations with the MAGIC telescopes of 3 sources, a symbiotic nova YY Her, a dwarf nova ASASSN-13ax and a classical nova V339 Del shortly after their outbursts. 
\end{abstract}

\maketitle

\thispagestyle{fancy}


\section{INTRODUCTION}
A classical nova is a thermonuclear runaway leading to the explosive ejection of the envelope accreted onto a white dwarf (WD) in a binary system in which the companion is either filling or nearly filling its Roche surface \citep{be08,st12,wr14}.
They are a type of cataclysmic variables, i.e. optically variable binary systems with a mass transfer from a companion star to WD.
Novae are typically detected first in optical observations when the brightness of the object increases by 7-16 magnitudes. 
The energy spectra of novae often contain a thermal X-ray continuum. 
The symbiotic novae, like the classical novae, are also initiated by a thermonuclear explosion on the surface of the WD.
However in the case of symbiotic novae, the WD is deep immersed in the wind of a late-type companion star (see e.g. \cite{sh11,sh12}).

The diffusive shock acceleration at the blast wave of symbiotic novae was expected to accelerate particles up to energies of a few TeVs \cite{th07}.
In 2010 the first GeV gamma-ray emission was detected by the \fermilat\ from the symbiotic nova V407 Cyg \cite{ab10}.
Such gamma-ray emission can be explained in terms of either leptonic or hadronic models \cite{ab10,novaescience}. 
In the former case, the local radiation fields create a target for the inverse Compton (IC) scattering of the electrons.
On the other hand, protons accelerated in the same conditions can interact with the matter producing gamma-rays via proton-proton interactions.
Several models have been put forward to explain the GeV radiation. 
For instance, the GeV gamma-ray emission can be attributed to the IC process on the strong radiation field of the red giant \cite{sb12}.
The same model predicts a second component in the TeV range due to proton-proton interactions with the wind of the red giant.
Also \cite{md13} consider acceleration of leptons and hadrons in the nova shock.
In that model the magnetic field, which determines the acceleration efficiency, is obtained assuming an equipartition with the thermal energy density upstream of the shock.
The GeV $\gamma$-ray emission is then a product of IC scattering of the nova light by the electrons.

In the last few years the \fermilat\ has discovered GeV gamma-ray emission from a few more novae: V1324 Sco, V959 Mon, V339 Del, and V1369 Cen \cite{cjs13, novaescience}. %% , and V745 Sco ,cjs14
Most of these sources are classical novae. 
Contrary to the symbiotic ones, they do not exhibit a strong wind of the companion star.
Interestingly, symbiotic and classical novae all exhibit similar spectral properties of the GeV emission.  
In classical novae the particles acceleration can occur e.g. on a bow shock between the nova ejecta and the interstellar medium or in weaker internal shocks due to inhomogenuity of the nova ejecta \cite{novaescience}.
In particular orbital motion of the system can lead to shaping the nova ejecta into a combination of a faster polar wind of the WD ejecta, and a denser material drifted along the equatorial plane \cite{cho14}.

So far no very-high-energy (VHE; E$>$100 GeV) gamma-ray emission has been detected from any nova event. 
VERITAS observations of V407 Cyg which started 10 days after the nova explosion yielded a differential upper limit on the flux at 1.6 TeV of $2.3 \times 10^{-12}\, \mathrm{erg\,cm^{-2}\,s^{-1}}$\cite{al12}
 
Beginning in Fall 2012 the MAGIC telescopes conducted a nova follow-up program in order to detect a possible VHE gamma-ray component. 
The program was first aimed on symbiotic novae. 
After the reports of detection of GeV emission from classical novae by the \fermilat, the program was extended also to bright classical novae and opened up to additional outbursts from cataclysmic variables. 

In here we report on MAGIC and \fermilat\ (see Section~\ref{sec:ins} for description of the used instruments and analysis methods) observations of the 3 sources observed within this program: a symbiotic nova YY Her (Section~\ref{sec:yyher}), a dwarf nova ASASSN-13ax (Section~\ref{sec:asassn}) and a classical nova V339 Del (Section~\ref{sec:v339del}).

\section{Instruments} \label{sec:ins}
The three sources were first detected and observed by optical instruments. 
The results of the MAGIC observations were supported by the analysis of quasi-simultaneous \fermilat\ observations. 

\subsection{MAGIC telescopes}
The VHE gamma-ray observations were performed with the MAGIC telescopes. 
MAGIC is a system of two 17\,m Cherenkov telescopes located on the Canary Island of La Palma at a height of 2200 m a.s.l.
The telescopes can perform observations of gamma rays with energies as low as $\sim$50\,GeV and up to tens of TeV. 
During Summer 2011 and 2012 MAGIC underwent a major upgrade \cite{mup1}. 
After the upgrade the sensitivity of the MAGIC telescopes in the best energy range ($\gtrsim300\,$GeV) is $\sim 0.6\%$ of Crab Nebula flux in 50\,h of observations \cite{mup2}.
All the data used for this paper were taken after the upgrade.
The data were analyzed using the standard analysis chain \cite{magic_mars, mup2}.
The significance of a gamma-ray excess was computed according to Eq.~17 of \cite{lm83}.
The upper limits on the flux were calculated following the approach of \cite{ro05} using 95\% C.L. and accounting for a possible 30\% systematic uncertainty on the effective area of the instrument. 

\subsection{\fermilat}
The \fermilat , launched in 2008, is a pair-conversion telescope that detects photons with energies from $20\,$MeV to $>300$\,GeV \cite{at09}.
Thanks to a large field of view ($\sim 2.4$ sr), the \fermilat\ observatory, operated in scanning mode, provides coverage of the full sky every three hours enabling searches for transient sources and overlap with ground-based observatories.
We analyzed the LAT data in the energy range 100 MeV $-$ 300 GeV using an unbinned maximum likelihood method \cite{mat96} as implemented in the {\it Fermi} Science Tools v9r32p5, the P7REP\_SOURCE\_V15 LAT Instrument Response Functions (IRFs), and associated standard Galactic and isotropic diffuse emission models\footnote{The P7REP data, IRFs, and diffuse models (gll\_iem\_v05.fit and iso\_source\_v05.txt) are available at http://fermi.gsfc.nasa.gov/ssc.}.
We selected events within a region of interest (ROI) of $15^\circ$ centered on the LAT best position (following \cite{novaescience}) for V339 Del and required a maximum zenith angle of $100^\circ$ in order to avoid contamination from Earth limb photons. 
Additionally, we applied a gtmktime filter (no.3) recommended for combined survey and pointed mode observations\footnote{http://fermi.gsfc.nasa.gov/ssc/data/analysis/\\documentation/Cicerone/Cicerone\_Likelihood/\\Exposure.html}, selecting good quality data at times when either the rocking angle was less than $52^\circ$ or the edge of the analysis region did not exceed the maximum zenith angle at $100^\circ$.
Sources from the 2FGL catalogue \cite{nol12} located within the ROI were included in the model used to perform the fitting procedure. 

\section{Sources observed}
We report here results of the MAGIC and \fermilat\ observations of YY Her, ASASSN-13ax and V339 Del. 

\subsection{YY Her}\label{sec:yyher}

\begin{figure}[t]
\includegraphics[width=0.49\textwidth]{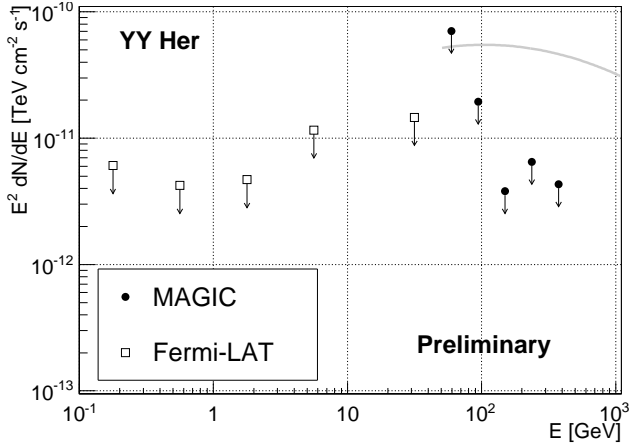}
\caption{Differential upper limits on the flux from YY Her as measured by the \fermilat\ (empty squares) and MAGIC (full circles). 
See text for details of the time ranges covered by the points.
For comparison a spectrum of Crab Nebula is shown with a gray curve. 
}\label{fig:yyher_ul}
\end{figure}

YY Her is a symbiotic nova system that undergoes a recurrent pattern of outbursts. 
MAGIC observations of YY Her occurred on the night of 2013 Apr 22\textsuperscript{nd}/23\textsuperscript{rd}, 7 days after the optical maximum. 
No significant VHE gamma-ray emission was detected. 
We computed flux upper limits at 95\% confidence level obtaining $<5.0\times \mathrm{10^{-12} ph\,cm^{-2}\,s^{-1}}$ above 300 GeV. 
Also in \fermilat\ no emission was detected over a longer interval 2013 Apr 10\textsuperscript{th} to Apr 30\textsuperscript{th} (MJD 56392.5 to 56412.5). 
Upper limits at 95\% confidence level were set as $2.8 \times 10^{-8} \mathrm{ph\,cm^{-2}\,s^{-1}}$ above 100\,MeV.
Differential upper limits obtained from the \fermilat\ and MAGIC observations of YY Her are shown in Fig.~\ref{fig:yyher_ul}.

\subsection{ASASSN-13ax}\label{sec:asassn}
\begin{figure}[t]
\includegraphics[width=0.49\textwidth]{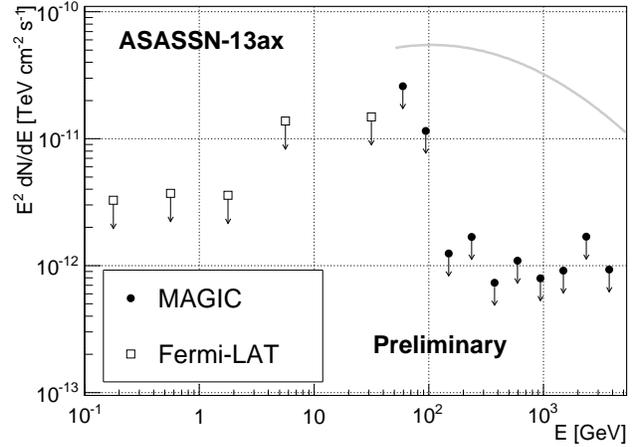}
\caption{Differential upper limits on the flux from ASASSN-13ax as measured by the \fermilat\ (empty squares) and MAGIC (full circles). 
See text for details of the time ranges covered by the points.
For comparison a spectrum of Crab Nebula is shown with a gray curve. 
}\label{fig:asassn_ul}
\end{figure}

ASASSN-13ax is a member of a different class of cataclysmic variables, the dwarf novae, which are known for significantly weaker optical outbursts (2-6 magnitudes) than classical novae. 
Instead of undergoing a thermonuclear explosion on the surface of the WD, these outbursts are caused by the gravitational energy release from a partial collapse of the accretion disk surrounding the WD.
The MAGIC observations were performed on two consecutive nights starting on 2013 Jul 4\textsuperscript{th}, soon after the optical outburst seen on 2013 Jul 1\textsuperscript{st}.
In the absence of detectable VHE emission, upper limits at 95\% confidence level were set as $<1.5\times \mathrm{10^{-12} ph\,cm^{-2}\,s^{-1}}$ above 300 GeV.
Emission was not detected in the LAT over the interval 2013 Jun 25\textsuperscript{th} to Jul 15\textsuperscript{th} (MJD 56468.5 to 56488.5). 
Upper limits at 95\% confidence level were set as $1.6 \times 10^{-8} \mathrm{ph\,cm^{-2}\,s^{-1}}$ above 100\,MeV. 
Differential upper limits obtained from the \fermilat\ and MAGIC observations of ASASSN-13ax are shown in Fig.~\ref{fig:asassn_ul}

\subsection{V339 Del}\label{sec:v339del}

\begin{figure}
\includegraphics[width=0.49\textwidth]{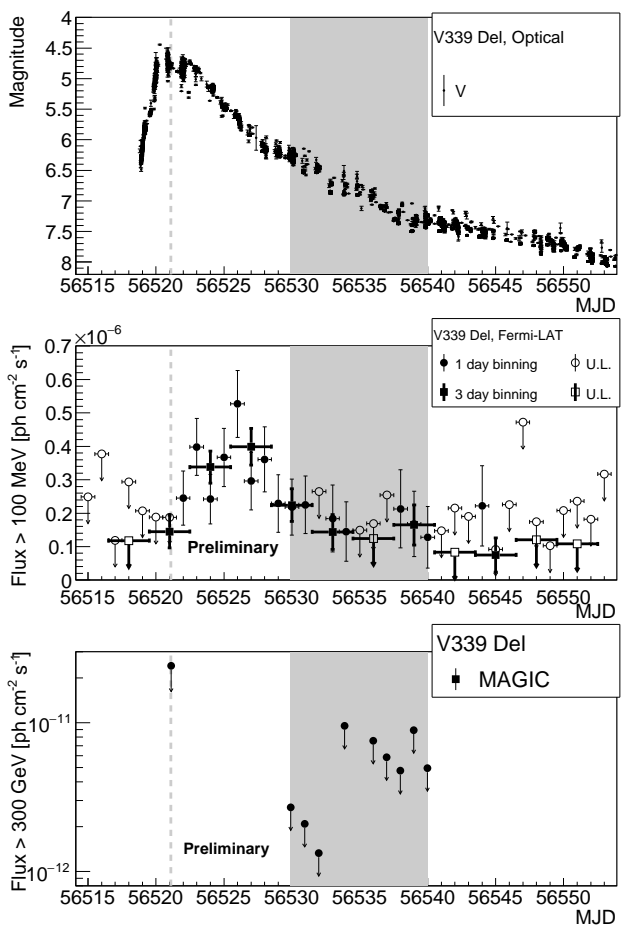}
\caption{
Multi-wavelength light curve of V339 Del during the outburst in August 2013.
Top panel: optical observations in the V band obtained from AAVSO-LCG\protect\footnote{http://www.aavso.org/lcg} service.
Middle panel: the \fermilat\ flux (filled symbols) and upper limits (empty symbols) above 100 MeV in 1-day (circles, thin lines) or 3-day (squares, thick lines).
A 95\% C.L. flux upper limit is shown for time bins with TS$<$4.
Bottom panel: Upper limit on the flux above 300 GeV observed with MAGIC telescopes.
The gray band shows the observation nights with MAGIC. 
The dashed gray line shows a MAGIC observation night affected by bad weather.}
\label{fig:del_mwl}
\end{figure}

V339 Del was a fast, classical nova detected by optical observations on 2013 Aug 16\textsuperscript{th}  (CBET \#3628). 
The nova was exceptionally bright reaching a magnitude of V$\sim 5\,$mag (see top panel of Fig.~\ref{fig:del_mwl}), and it triggered follow-up observations at frequencies ranging from radio to VHE gamma-rays.
Photometric measurements suggest a distance for V339 Del of $4.5\pm0.6$\,kpc \cite{sch14}.
The spectroscopic observations performed on MJD 56522.1 revealed emission wings extending to about $\pm 2000\,$km/s and a Balmer absorption component at a velocity of $600\pm 50\,$km/s \cite{sh13}.
The pre-outburst optical images revealed the progenitor of nova V339 Del to be a blue star \cite{de13}. 

\begin{figure}
\centering
\includegraphics[width=0.49\textwidth]{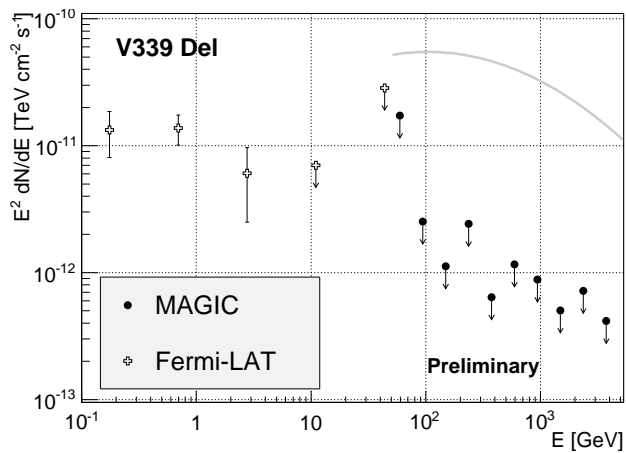}
\caption{
Differential upper limits on the flux from V339 Del as measured by MAGIC (filled circles) and the flux measured by \fermilat\ (empty crosses) in the same time period, 25\textsuperscript{th} of August and 4\textsuperscript{th} of September.
For comparison a spectrum of Crab Nebula is shown with a gray curve. 
}\label{fig:del2013_sed}
\end{figure}

Originally MAGIC observations of V339 Del were motivated by its extreme optical outburst.
Soon after MAGIC started observations they were additionally supported by the detection of GeV emission by the \fermilat\ from the direction of V339 Del.
The MAGIC observations started already on the night of 2013 Aug 16/17\textsuperscript{th}, however they were marred by bad weather conditions. 
The good quality data used for most of the analysis spanned 8 nights between 2013 Aug 25\textsuperscript{th} and Sep 4\textsuperscript{th}. 
The total effective time was 11.6\,h.
In addition to the nightly upper limits we performed a dedicated analysis of the poor quality (affected by calima, a dust layer originating from Sahara) night of 2013 Aug 16/17\textsuperscript{th}.
We applied an estimated energy and collection area corrections based on LIDAR measurements \cite{fr14}.
No VHE gamma-ray signal was found from the direction of V339 Del.
We computed a night by night integral upper limit above 300\,GeV (see bottom panel of Fig.~\ref{fig:del_mwl}. 
The differential upper limits for the whole good quality data set computed in bins of energy are shown in Fig.~\ref{fig:del2013_sed}.

Nova V339 Del was the subject of a \textit{Fermi} Target of Opportunity (ToO) observation \cite{AT5302} triggered by the optical discovery (CBET \#3628); the ToO started on 2013 Aug 16\textsuperscript{th} and lasted for six days.
The gamma-ray emission from V339 Del was first detected by \fermilat\ in 1-day bins on Aug 18\textsuperscript{th} \cite{novaescience}. 
The emission peaked on Aug 22\textsuperscript{nd} and entered a slow decay phase afterwards (see middle panel in Fig.~\ref{fig:del_mwl}).
For the light-curves, the data were fit using a power law spectral model initially leaving the photon index and the normalization free to vary.
We then fixed the photon index at the average value of 2.3 calculated over the most significant detections (Test Statistic values TS$>$9)\footnote{The source significance is $\sim$ sqrt(TS) assuming one degree of freedom}. 
The LAT Spectral Energy Distribution (SED) of V339 Del shown in Fig.~\ref{fig:del2013_sed} was extracted in five logarithmically spaced energy bins from 100 MeV to 100 GeV.
Similarly to the light-curves, energy binned data shown in Fig.~\ref{fig:del2013_sed} were fit using a simple power law and showing a 95\% C.L. upper limit for bins with TS$<$9. 
In the period coincident with the MAGIC observations (2013 Aug 25\textsuperscript{th} to Sep 4\textsuperscript{th}) the \fermilat\ spectrum can be described by an effective power law with an index of $2.37\pm0.17$ and flux above 100 MeV of $(0.15\pm 0.04) \times 10^{-6} \mathrm{ph\,cm^{-2}\,s^{-1}}$. 
The rather low statistical significance (TS=49) does not constrain the value of an exponential cut-off of the emission in this period.
Note, however, that the most energetic photon, with $E=5.9$\,GeV was recorded on Aug 30\textsuperscript{th}, i.e. within the time period covered by the MAGIC observations.
The \fermilat\ analysis for a broader time range, 2013 Aug 22\textsuperscript{nd} to Sep 12\textsuperscript{th} (MJD 56526-56547), covering the whole decay phase of the Fermi-LAT light curve allowed us to obtain a more significant signal with a TS of 121. 
Nevertheless  we obtain a similar value of flux above 100 MeV,  $(0.13\pm 0.03) \times 10^{-6} \mathrm{ph\,cm^{-2}\,s^{-1}}$, for this broader period.
The spectrum in this case can be described, with improved significance of $3.3\sigma$ with respect to the simple power law, by an exponentially cut-off power law with an index of $1.44\pm 0.29$ and a cut-off energy of $1.6\pm0.8$\,GeV. 

\section{Conclusions}\label{sec:conc}
The MAGIC telescopes performed observations of 3 objects: the symbiotic nova YY Her, the dwarf nova ASASSN-13ax and the classical nova V339 Del. 
No significant VHE gamma-ray emission was found from the direction of any of them.
Out of these three objects, V339 Del is the only one detected at GeV energies.
It has also extensive optical observations which shed some light on both the companion star and the photosphere of the nova.
Therefore it has the highest potential for constraining the leptonic and hadronic processes in novae.
MAGIC will continue follow-up observations of the promising novae candidates in the following years.

\bigskip % extra skip inserted
\begin{acknowledgments}
\small
We would like to thank
the Instituto de Astrof\'{\i}sica de Canarias
for the excellent working conditions
at the Observatorio del Roque de los Muchachos in La Palma.
The support of the German BMBF and MPG,
the Italian INFN, 
the Swiss National Fund SNF,
and the ERDF funds under the Spanish MINECO
is gratefully acknowledged.
This work was also supported
by the CPAN CSD2007-00042 and MultiDark CSD2009-00064 projects of the Spanish Consolider-Ingenio 2010 programme,
by grant 127740 of the Academy of Finland,
by the Croatian Science Foundation (HrZZ) Project 09/176 and the University of Rijeka Project 13.12.1.3.02,
by the DFG Collaborative Research Centers SFB823/C4 and SFB876/C3,
and by the Polish MNiSzW grants 745/N-HESS-MAGIC/2010/0 
and NCN 2011/01/B/ST9/00411. %% WB & JS

The \textit{Fermi} LAT Collaboration acknowledges generous ongoing support
from a number of agencies and institutes that have supported both the
development and the operation of the LAT as well as scientific data analysis.
These include the National Aeronautics and Space Administration and the
Department of Energy in the United States, the Commissariat \`a l'Energie Atomique
and the Centre National de la Recherche Scientifique / Institut National de Physique
Nucl\'eaire et de Physique des Particules in France, the Agenzia Spaziale Italiana
and the Istituto Nazionale di Fisica Nucleare in Italy, the Ministry of Education,
Culture, Sports, Science and Technology (MEXT), High Energy Accelerator Research
Organization (KEK) and Japan Aerospace Exploration Agency (JAXA) in Japan, and
the K.~A.~Wallenberg Foundation, the Swedish Research Council and the
Swedish National Space Board in Sweden.
Additional support for science analysis during the operations phase is gratefully acknowledged
from the Istituto Nazionale di Astrofisica in Italy and the Centre National d'\'Etudes Spatiales in France.
\end{acknowledgments}

\bigskip % extra skip inserted
%\begin{thebibliography}{9}   % Use for  1-9  references

\end{document}